\documentclass[journal=jctcce,manuscript=article]{achemso}

\usepackage[T1]{fontenc} 
\usepackage{amsmath}
\newcommand*\bmat[1]{\mathbf{#1}}
\usepackage[labelfont={bf}]{caption}
\usepackage{mathtools}
\newtagform{eqt}{(S}{)}

\author{Michelle A. Hunter}
\author{Baris Demir}
\author{Charlotte F. Petersen}
\altaffiliation{Current address: School of Chemistry, University of Sydney, Sydney, NSW, Australia}
\author{Debra J. Searles}
\email{d.bernhardt@uq.edu.au}
\phone{+61 (0)7 33463939}
\affiliation[The University of Queensland AIBN]
{Australian Institute for Bioengineering and Nanotechnology, The University of Queensland, Brisbane, Qld 4072, Australia}
\alsoaffiliation[The University of Queensland SCMB]
{School of Chemistry and Molecular Biosciences, The University of Queensland, Brisbane, Qld 4072, Australia}

\title[Local self-diffusion coefficients]
  {A new framework for computing a general local self-diffusion coefficient using statistical mechanics}

\begin{document}

\begin{abstract}
  Widely applicable, modified Green-Kubo expressions for the local diffusion coefficient ($D_l$) are obtained using linear response theory. In contrast to past definitions in use, these expressions are statistical mechanical results. Molecular simulations of systems with anisotropic diffusion and an inhomogeneous density profile confirm the validity of the results.  Diffusion coefficients determined from different expressions in terms of currents and velocity correlations agree in the limit of large systems.  Furthermore, they apply to arbitrarily small local regions, making them readily applicable to nanoscale and inhomogeneous systems where knowledge of $D_l$ is important.
\end{abstract}

\section{Introduction}

The accurate determination of local transport phenomena is of great importance to chemical science. This is because local behavior in a heterogeneous system may be significantly different to the behavior in a bulk system, meaning it is dependent on spatial position. Consequently, position-dependent diffusivity is very important in a wide variety of applications, including the design of energy storage devices, where ionic conductivity through separators and at interfaces is relevant, in heterogeneously catalysed reactions where surface diffusion differs from the bulk, and in consideration of transport mechanisms in biological cells. The approaches outlined in this paper can be applied to study of different diffusion coefficients, but here we focus on self-diffusion coefficients.  

The characterisation of local properties in inhomogeneous systems requires a larger system to be divided into local regions of interest. Whilst expressions for the local viscosity and thermal conductivity have been determined and applied,\cite{Todd1995, Hansen2007,Galliero2012,Hoang2012} defining local diffusion coefficients has posed particular problems as the diffusing particles will move between different local regions, creating ambiguity as to which particles should be chosen as belonging to a local region as the particles move. In molecular dynamics (MD) simulations, the bulk self-diffusion coefficients of inhomogeneous and confined liquid systems are usually calculated using the mean-squared displacement (MSD) of the particles with time, or from the integral of the velocity autocorrelation function (VACF). However, directly applying the MSD or VACF expressions to small sections of a system is generally not possible because particles move between regions in the period of measurement of the MSD, or the time for convergence of the integral of the VACF, requiring some choice of the observed time frame, or criteria to define particles belonging to the local region. Consequently, different results will be obtained depending on which particle trajectories are included in the ensemble average.

There have been various modified Green-Kubo relations proposed to measure local diffusion coefficients.\cite{Buhn2004,Buhn2006a,Braga2014,Chilukoti2016} For example, Buhn et al.\cite{Buhn2004a,Buhn2006a} define a local diffusion coefficient by including the velocity correlations of the particles that start in the local region at the time origin, which they justify based on number of particles moving outside the local region within the correlation time as being negligible. Another example is in Chilukoti et al.,\cite{Chilukoti2016} where they defined the local self-diffusion coefficients of layers of liquid-vapor interfaces by integrating velocity correlations of molecules that stayed continuously within the layer for the whole time period observed. A third example is in Braga et al.,\cite{Braga2014} where they define the diffusion coefficient by including correlations of particles that are in the local region at the initial time and at the time of interest, but do not require the molecules to stay within the layer the whole time. 

There have also been examples of modified calculations of MSDs to calculate local self-diffusion coefficients which have largely been applied in the directions parallel to an interface.\cite{Buhn2004a,Liu2004, Chilukoti2015, Colmenares2009, MercierFranco2016,Benjamin1992,Chio2020,Fabian2016,Fabian2020} However, an issue with applying a local MSD is that if the region is small, many of the molecules will leave the region before the mean-squared displacement reaches the linear regime, or that many of the particles will leave the local region in the measured time. To remedy this, others have used a survival probability, which effectively scales the value of the MSD based on the probability of a particle of staying within the local region. \cite{Liu2004, Chilukoti2015, Colmenares2009, MercierFranco2016}

There have also been examples of local self-diffusion coefficients using methods other than the Green-Kubo and Einstein expressions.\cite{Liu2004,Zaragoza2019,Olivares-Rivas2013,Vermorel2017,MercierFranco2016} Liu et al. developed a dual simulation method which combines molecular dynamics and Langevin dynamics by fitting the calculated Langevin survival probability to the MD calculated survival probability to determine a diffusion coefficient near an interface.\cite{Liu2004} Another example in Zaragoza et al.\cite{Zaragoza2019} uses a modified Stokes-Einstein equation to calculate the viscosity of confined systems and to calculate the local diffusion coefficients in a carbon slit and nanotube.\cite{Zaragoza2019}
 
As shown above, there have been many methods have been proposed to measure local-self diffusion coefficients, which have required some choice of the particles included, direction of diffusion, or choice of boundary conditions for the measurement of the diffusion coefficient. It is therefore desirable to have a clearly defined statistical mechanical result which is easily implementable in molecular dynamics simulations. This is the goal of this manuscript. 

The Green-Kubo (GK) expression for the self-diffusion coefficient, $D$, of a bulk liquid is given by the long time integral of correlations in the particle velocities:
\begin{equation} \label{globalGK}
	\begin{split}
		D = \lim\limits_{t \to \infty} \frac{1}{3N} \int^t_0 ds \langle \mathbf{v}(0) \cdot \mathbf{v}(s) \rangle, 
	\end{split}
\end{equation}
where $N$ is the number of particles, $\langle ... \rangle$ denotes an ensemble average and $\mathbf{v}$ is the vector of velocities of all the particles. It has been shown that the GK relation can be obtained from a more general theorem for nonequilibrium systems, called the dissipation theorem.\cite{Evans2008} 
In this manuscript we derive a GK-like expression for the local self-diffusion coefficient from the dissipation theorem, and discover that correlations between local and global quantities must be considered.
We rigorously test our expression in molecular dynamics simulations of homogeneous systems, where the local self-diffusion coefficient, $D_{l}$, is expected to agree with the self-diffusion coefficient of the whole system, and confirm that our result is independent of the shape and size of the local region compared to the whole system. We then calculate $D_{l}$ in a single-component liquid with an inhomogeneous density, and show that its values are reasonable in comparison to homogeneous systems of equivalent average densities.

\section{Theoretical Background: Green-Kubo Relations and the Dissipation Theorem}
The Green-Kubo relations enable linear transport coefficients to be measured through correlation functions at equilibrium. In this section, the relationship between the dissipation theorem and the GK relations that has already been established\cite{Evans2008}  will be summarized. 

Consider an $N$-particle system with nonequilibrium dynamics described by the following `color conductivity' equations of motion:\cite{Evans2008b} 
\begin{equation}
	\label{equationsofmotion}
	\begin{split}
		\mathbf{\dot{q}}_{i} &= \frac{\mathbf{{p}}_{i}}{m}\\
		{\mathbf{\dot{p}}_{i}} &= {\mathbf{F}_{I,i}}+ c_{i}\mathbf{F}_{c} - \eta \mathbf{p}_{i}   \\             
		\dot{\eta}  &= \frac{1}{Q}\bigg (\frac{\sum\nolimits_{i} \mathbf{p}_{i} \cdot \mathbf{p}_{i}}{m}-\frac{3N-3}{\beta} \bigg  ),
	\end{split}
\end{equation}
where $m$ is the particle mass (assumed here to be the same for all particles for simplicity, but this is readily generalized), $\mathbf{F}_{I,i}$ is the total interparticle force on particle $i$, $\mathbf{F}_{c}$ is the color field which drives the flow in the system, and $c_{i}$ is the color charge coupling particles to the color field.  The term $\eta$ is a thermostatting variable for the Nos\'e-Hoover thermostat, $Q$ is the mass of the thermostat, $\beta$ = $1/k_{B}T$ and $k_{B}$ is the Boltzmann constant. 

In earlier work\cite{Evans2008} it was shown how the GK relations can be obtained from a special case of the transient-time correlation function (TTCF) formalism, which is part of a more general theorem for nonequilibrium systems, called the dissipation theorem (DT). The DT demonstrates that the ensemble average of an arbitrary phase function, $B(\mathbf{\Gamma}(t))$, can be related to its initial value $\langle{B(\mathbf{\Gamma}(0))\rangle}$ and the integral of its correlation with the dissipation function, $\Omega{(\mathbf{\Gamma}(t))}$:\cite{Evans2008b}
\begin{equation}\label{general_phase}
	\langle{B(\mathbf{\Gamma}(t))\rangle} = \langle{B(\mathbf{\Gamma}(0))\rangle} + \int^t_0 ds \langle{\Omega(\mathbf{\Gamma}(0))B(\mathbf\Gamma(s))\rangle}, 
\end{equation}
where ${\mathbf{\Gamma} \equiv \{\mathbf{\mathbf{\Gamma}}_{1},\mathbf{\mathbf{\Gamma}}_{2},...\mathbf{\mathbf{\Gamma}}_{N}\}\equiv\{\mathbf{p}_{1},\mathbf{q}_{1},\mathbf{p}_{2},\mathbf{q}_{2}.....\mathbf{p}_{N},\mathbf{q}_{N}\}}$ is the phase space vector. Here, the vectors for $\mathbf{q}_{i}$ and $\mathbf{p}_{i}$ represent the positions and momenta of the $i$th particle. Note that the ensemble average is with respect to the initial distribution function.

For a system starting in canonical equilibrium, the dissipation function, $\Omega{(\mathbf{\Gamma}(t))}$, can be defined as:\cite{Sevick2008a}
\begin{equation}\label{dissipation}
	\Omega{(\mathbf{\Gamma}(t))} = \frac{dH(\mathbf{\Gamma}(t))}{dt} - \Lambda(\mathbf{\Gamma}(t)),
\end{equation}
where \textit{H} is internal energy of the system, and $\Lambda (\mathbf{\Gamma}(t))$ is phase space compressibility, which is zero for Hamiltonian dynamics. The dissipation function is zero at all phase points in an equilibrium system.
For the system considered here, the dissipation function can be written in terms of the color current, $\mathbf{J}(t)$:\cite{Searles2000b}
\begin{equation}
	\Omega(\mathbf{\mathbf{\Gamma}}(t)) = N \beta \mathbf{J}(t) \cdot \mathbf{F}_{c},
	\label{coldissipation}
\end{equation}
where the color current is defined as 
\begin{equation}\label{colorcurrent}
	\mathbf{J}(t) = \frac{1}{N} \sum\nolimits_{i} c_{i} \mathbf{v}_{i}.
\end{equation}
Here $\mathbf{v}_{i}$ denotes the velocity of particle $i$ in the direction of the applied color field, and we set $c_{i}= (-1)^i$ in this work. 

If the field is applied in one direction, $\alpha$ ($\alpha = x, y$ or $z$), with magnitude $F_{c}$, then substituting equation (\ref{coldissipation}) into equation (\ref{general_phase}) for the phase variable $B=J_{\alpha}$, we obtain:
\begin{equation}\label{flux}
	\langle {J_{\alpha}}(t) \rangle  = N \beta {F}_{c} \int^t_0 ds \langle {J_{\alpha}}(0) {J_{\alpha}}(s) \rangle,
\end{equation}
\noindent where $t$ is the time after application of the color field to the equilibrium system. Equation 
(\ref{flux}) is the TTCF expression for the color current.\cite{Evans1994a} In limit $F_c \to 0$, the ensemble average on the right hand side becomes the ensemble average of the equilibrium system (\textit{i.e.} both the ensemble and the dynamics are equilibrium). As shown in Section 1.1 of the Supporting Information,  in the large system limit,  $ N \langle {J_{\alpha}}(0) {J_{\alpha}}(t) \rangle $ =  $\langle {v_{\alpha,i}}(0){v_{\alpha,i}}(t) \rangle $,\cite{Evans2008b} so the results obtained from the color current autocorrelation function and the velocity autocorrelation function become equal. Then 
\begin{equation}\label{Dcolorcurrent}
	\begin{split}
		\lim_{{F}_{c}\to 0} \lim\limits_{t \to \infty} \frac{\langle {J_{\alpha}} \rangle}{\beta {F}_{c}} = N \int^\infty_0 ds \langle {J_{\alpha}}(0){J_{\alpha}}(s) \rangle_{\text{eq}} \\  \overset{\mathrm{N \to \infty}}{=}  \frac{1}{N}\sum_{i=1}^N\int^\infty_0 ds \langle {v_{\alpha,i}}(0) {v_{\alpha,i}}(s) \rangle_{\text{eq}} = D_{\alpha}.
	\end{split}
\end{equation}
This demonstrates that $D_{\alpha}$ can be determined from the correlations at equilibrium, and also from the nonequilibrium linear response of the color current.

\section{Local Self-Diffusion Coefficient}
Since the DT is true for arbitrary phase functions it can be rigorously applied to a local phase function $B_{l}$, that is only dependent on $\mathbf{\Gamma}_i$  for  particles $i$ in the local region, to give a local TTCF expression. It has been applied to simple bulk and confined liquid systems with an applied color field to measure the local color current,\cite{Talaei2012,Brookes2016} and provides a means to determine changes in a phase variable for a local region of a system:
\begin{equation}
	\label{localTTCF}
	\langle{B_{l}(\mathbf{\Gamma}(t))\rangle} = \langle{B_{l}(\mathbf{\Gamma}(0))\rangle} + \int^t_0 ds \langle{\Omega(\mathbf{\Gamma}(0))B_{l}(\mathbf{\Gamma}(s))\rangle}.
\end{equation}

Since $\Omega(\mathbf{\Gamma}(0))$ depends on $\mathbf{\Gamma}_i$ of all particles (\textit{i.e.} it is a global function) it was revealed that if the current was determined using purely local correlations, the values obtained would depend on the size of the local region and the correlation length.\cite{Talaei2012} This result is important for the calculation of local properties, as it quantifies the error associated with excluding the correlations with particles outside the local region when determining the local properties.

We define a local color current:
\begin{equation}\label{localcolorcurrent}
	\mathbf{J}_{l}(t) = \frac{1}{N_{l}} \sum\nolimits_{i} c_{i} \mathbf{v}_{i}(t) S({\mathbf{q}}_i(t)),
\end{equation}
where $S({\mathbf{q}}_i(t))$ is a switch which is equal to 1 if particle $i$ is in the local region $l$, and 0 if the particle is outside $l$, $N_{l} = \langle N_{l}(t) \rangle $ and is the ensemble average number of particles in $l$ at time $t$.
Substituting $\mathbf{J}_{l}(t)$ for $B_{l}(\mathbf{\Gamma}(t))$ into equation (\ref{localTTCF}) and using (\ref{coldissipation}) gives, 
\begin{equation}
	\langle{{J}_{\alpha,l}(t)\rangle} = \langle{{J}_{\alpha,l}(0)\rangle} + N \beta {F}_{c} \int^t_0 ds \langle {J_{\alpha}}(0) {J}_{\alpha,l}(s) \rangle, 
\end{equation}
\noindent where $\langle{{J}_{\alpha,l}(0)\rangle}$ = 0 because the system is at equilibrium and we obtain an expression for the response of the local color current in terms of a `global-local' correlation function. Using the same reasoning used to give equation (\ref{Dcolorcurrent}), a local self-diffusion coefficient, $D_{l}$, can be defined. This assumes that $D_{l}$ has the same dependence on the response of the local color current as $D$ has on the global current. Then, 
\begin{eqnarray}
	D_{\alpha,l} &\equiv& \lim_{{F}_{c}\to 0} \lim_{t\to \infty} \frac{\langle {J}_{\alpha,l}(t) \rangle}{\beta {F}_{c}} = N\int^\infty_0 ds \langle {J}_{\alpha}(0) {J}_{\alpha,l}(s) \rangle_{\text{eq}}\nonumber\\ &\overset{\mathrm{N \to \infty}}{=}&\frac{1}{N}\sum_{i=1}^N \int^\infty_0 ds \langle {v}_{\alpha,i}(0) {v}_{\alpha,i,l}(s) \rangle_{\text{eq}}\nonumber\\
	&=& \frac{1}{N}\sum_{i=1}^N \int^\infty_0 ds \langle {v}_{\alpha,i,l}(0) {v}_{\alpha,i}(s) \rangle_{\text{eq}},\label{localD}
\end{eqnarray}
where $ \langle {v}_{\alpha,i}(0) {v}_{\alpha,i,l}(s) \rangle  \equiv N/N_l \langle {v}_{\alpha,i}(0) {v}_{\alpha,i}(s)S({\mathbf{q}}_i(s)) \rangle $. The second equality in (\ref{localD}) is derived in Section 1.1 of the Supporting Information, where it is shown that although it is only valid in the large $N$ limit, it is not necessary for $N_l$ to be large. The last equality is derived in Section 1.3 of the Supporting Information and shows that the switch may be applied at the initial time, so that the correlation function contains contributions from the same set of particles for all times. Because we focus on the self-diffusion coefficient exclusively in this paper, we subsequently refer to it as the diffusion coefficient; and we refer to the diffusion coefficient of the whole system as the `global diffusion coefficient'. While (\ref{localD}) shows that the local diffusion coefficient is a function of the correlations between the local and global color currents, this expression fortunately reduces to the velocity correlation function of particles chosen to be in some local region at the initial time, which can be calculated in practice without full knowledge of the whole system. 

It is important to note here that though a local region must be defined to apply the equations in  (\ref{localD}), that the behaviour of particles outside the local region is included in the global terms in each of the equalities. That is, the correlations of particles within the local region and outside it are taken into account and the particles are able to freely leave the region. By this definition, this implies that there is always some nonlocal dependence of the local self-diffusion coefficient which will depend on the correlation time of the particles in question. If particles were unable to leave, they would not be able to travel far enough to achieve diffusive behaviour. 

\section{Simulation Methods}
In this work, we carry out equilibrium simulations using an in-house nonequilibrium MD code. Equilibrium simulations were carried out in the canonical ensemble to evaluate global and local diffusion coefficients defined by (\ref{Dcolorcurrent}) and (\ref{localD}). The equations of motion are given by (\ref{equationsofmotion}) with $\mathbf{F}_c=\mathbf{0}$. Interparticle interactions are treated with a 12-6 Lennard-Jones potential with a cutoff of $r_{\text{cut}} = 2.5$, so the value of the interaction potential is 0 for $r_{ij} > r_{\text{cut}}$.  Throughout this manuscript, Lennard-Jones reduced units are used with $\epsilon$, $\sigma$ and $m$ = 1. The temperature is $T$=1.0 in all cases, the particle density is $\rho$ = 0.80 for the homogeneous liquid in the cubic and non-cubic simulation cells where the number of particles, $N$, is 2048 for the cubic system and $N$ is 4096 for the non-cubic system.

In a separate section of this manuscript, we simulate an inhomogeneous system. To create a single-component inhomogeneous system, a sinusoidal force was applied in the $x$ direction of the cubic simulation cell of length $L_x =14.30$, by adding a term $ \text{sin}(2\pi x/L_x$) to the momentum equation of motion in equation (\ref{equationsofmotion}), resulting in a non-uniform density profile, where the average particle density is $\rho$ = 0.70 and $N$= 2048. 

A Nos\'e-Hoover thermostat of mass $Q=50$ is used to sample the canonical ensemble and a 4th order Runge-Kutta scheme is employed to integrate the equations of motion with a time step $dt$ = 0.002. Periodic boundary conditions are applied in the three Cartesian directions. All liquids are initialized with the particles on face-centered cubic lattices and then equilibrated for 50,000 timesteps. New time origins for the correlation functions are selected every 500 timesteps thereafter and a time-reversal mapping is used to obtain two starting phase points for the trajectories to ensure the numerical evaluation of $ \langle {v}_{\alpha,i}(0) \rangle $ is $0$.

\section{Demonstration of the Local Diffusion Coefficient}
\subsection{Homogeneous System with Isotropic Diffusion}
\begin{figure}[httb!]
    \centering
    \includegraphics[scale=0.5]{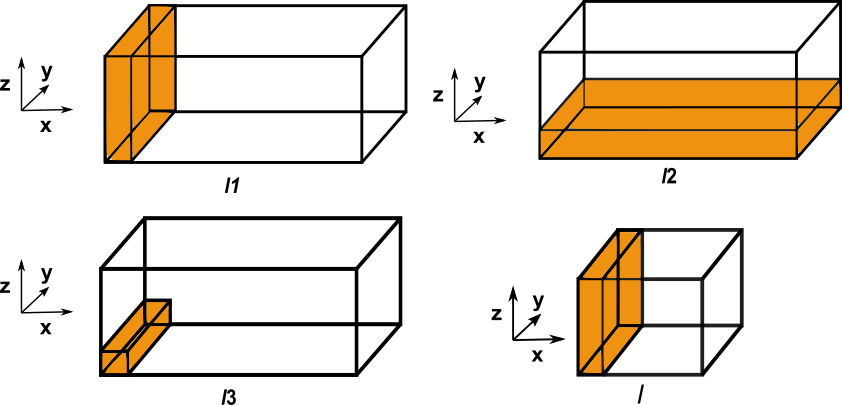}
    \caption{The systems and local regions that were investigated in this work. For all systems, $\rho = 0.80$. For the cubic system, $N$ = 2048. The dimensions of the cubic system were $x = y = z =$ 13.68 (shown bottom right). The dimensions of the region $l$ in the cubic system is $x$ = 1.00, and $y = z =$ 13.68. For the anisotropic system, N = 4096, $z/x$ = $y /x$ = 0.5 and $x = 27.36 $, $y = z = 13.68 $. Pictured are the three local regions which were explored in this study, $l1$, $l2$, and $l3$. Dimensions of $l1$ are $ x= 2.00 $, $y = z = 13.68 $, of $l2$ are $x = 27.36 $, $y = 13.68 $,  $z = 2.00$, and $l3$ are $x = z = 2.00$, $y=13.68$. }
    \label{fig:boxes}
\end{figure}

\begin{figure}[htbp!]   
	\centering
	\includegraphics[scale=0.50]{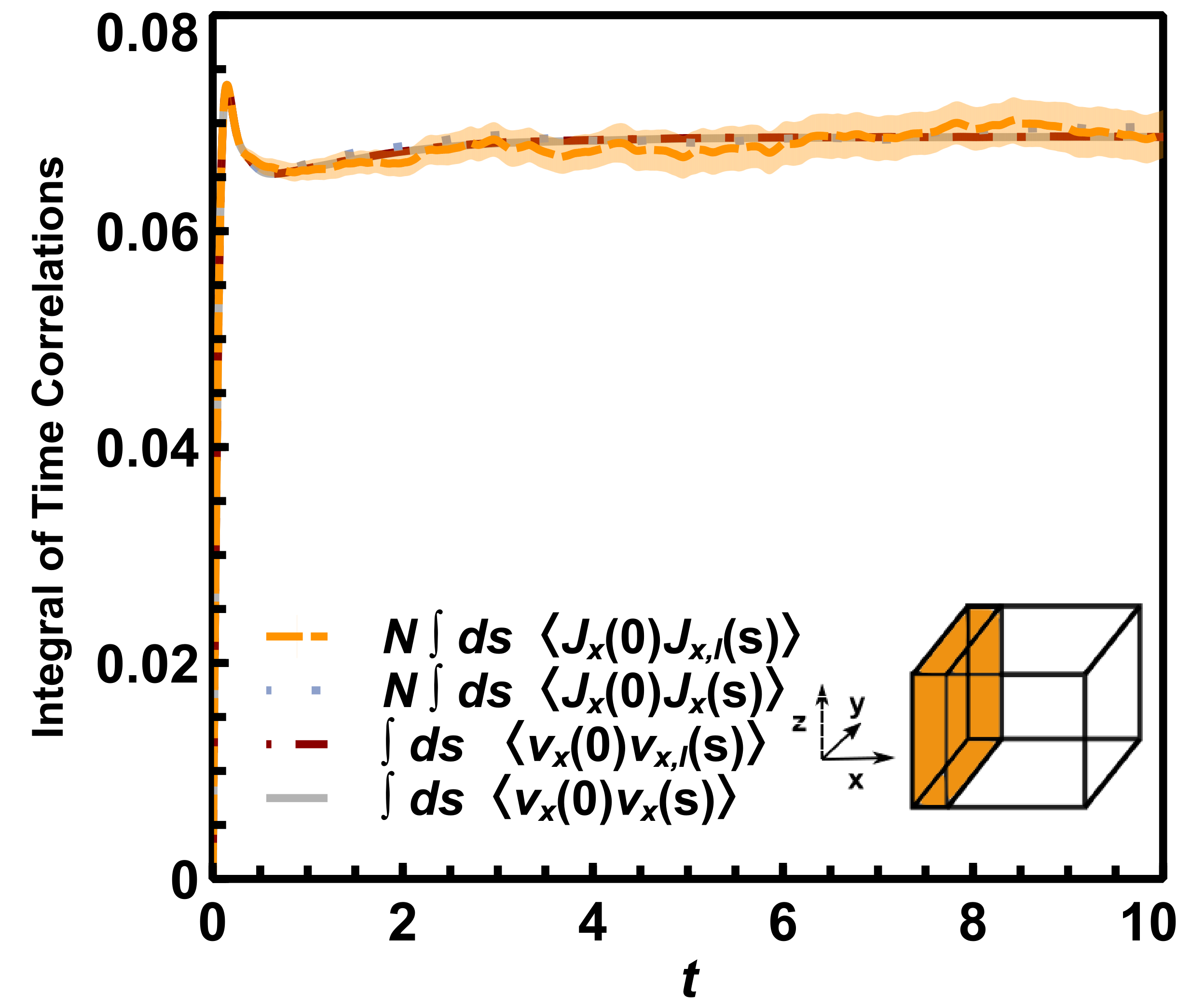}
	\caption{Integrals of time correlations of global color current  and global velocities from 0 to $t$, compared with the corresponding global-local correlations for the region $l$ in the $x$ direction. These quantities give $D_{x}$ and $D_{l,x}$ in the long time limit as per equations (\ref{Dcolorcurrent}) and (\ref{localD}). The error bars for the local color current correlation functions, shown in light orange, are one standard error in the mean.}
	\label{fig:GKTTCFcube}
\end{figure}

A homogeneous liquid system was simulated in a cubic simulation box (length 13.68) (see Figure \ref{fig:boxes}). In this case, the value of $D_l$ is expected to be equal to the global bulk diffusion coefficient for a box of this size, irrespective of the size and shape of the local region. A local region, $l$, was defined with dimensions $x = 1$ and $y = z = 13.68$.  An illustration of the local region and the system is shown in the inset of Figure~\ref{fig:GKTTCFcube}. 

It can be seen that $D_{x,l}$ given by the local color current and local velocities agrees with $D_{x}$, as the integrals of the time correlations overlap with the global values within error bars, shown in Figure \ref{fig:GKTTCFcube}. This agreement is the same in all directions (see Figure S1 of the Supporting Information), which demonstrates that $D_{\alpha,l}$ in this case is independent of the dimensions of the local box, despite the $x$ dimension of this local region being much smaller than the $y$ and $z$ dimensions. Furthermore, it can be seen that although the results obtained for $D_{\alpha,l}$ from the correlations of the color current agree with the other expressions, they are much noisier than the velocity correlation results.  This is expected, as discussed in Section 1.2 of the Supporting Information.

\subsection{Homogeneous System with Anisotropic Diffusion}
For a system with an anisotropic simulation box it is expected that the global diffusion coefficient may be anisotropic due to finite-size effects.\cite{Yeh2004,Celebi2020} We simulated a square-cuboid periodic simulation cell, where $z/x=y/x =0.5$, and found that $D_{x} < D_{y} = D_{z}$. However, since this anisotropic system is homogeneous, it is expected that the local diffusion coefficients in any region should still match the global diffusion coefficient. It is important to note that the finite size effects that we refer to here  are those that apply to the whole simulation cell due to the interaction of the simulation with its periodic images.\cite{Yeh2004,Celebi2020,Jamali2020} 
They are a feature of all periodic simulations, and are not due to the finite size of local regions we consider when measuring the local diffusion coefficient. Although it is possible to correct for finite-size effects in the calculation of bulk diffusion coefficients, the corresponding correction for local diffusion coefficients is unknown, and so we present all diffusion coefficients uncorrected. 


We define three local regions, $l1, l2$ and $l3$ which occupy different proportions of the full simulation cell. We find that the calculated local diffusion coefficients are indeed independent of the size and shape of the local regions, but instead depend on the geometry of the full simulation box, $D_{\alpha,l}$ = $D_{\alpha}$ for all differently sized and shaped regions (see Table  \ref{GKTCFan} and Figures S2 and S3 of the Supporting Information for details of the calculation). The anisotropy of the global diffusion coefficients is reflected in the local values even for very small local regions, indicating that our local expression is ideal for measuring extremely localized or anisotropic behavior.

	\begin{table}
		\begin{center}
		\begin{tabular}{| *{7}{c|} }
			\hline
			Box\textsuperscript{\emph{a}}    & \multicolumn{3}{c|}{Dimensions}
			& \multicolumn{3}{c|}{$D_{l}$ ($10^{-2})$} \\
			\hline
			&   $x$  &   $y$ & $z$ &    $D_{x,l}$  &   $D_{y,l}$  &   $D_{z,l}$    \\
			\hline 
			global\textsuperscript{\emph{b}}   &   27.36 &   13.68  & 13.68 &     6.86  &   7.08  &   7.08     \\
			\hline 
			1   &   2.00  &   13.68  & 13.68 &    6.87(1)\textsuperscript{\emph{c}}  &   7.07(1)  &   7.08(1)  \\
			\hline
			2   &  27.36  &  13.68  & 2.00 &   6.86    &   7.08   &   7.08  \\
			\hline 
			3   &   2.00  &   2.00  & 13.68 &    6.87(1)    &   7.10(2)   &   7.09(1) \\
			\hline
		\end{tabular}
		\caption{Diffusion coefficients ($D_{l}$) in local regions $l1$- $l3$ of a liquid system with anisotropic diffusion, calculated using the velocity autocorrelations.  }
		\label{GKTCFan}
		\end{center}
		\footnotesize{\textsuperscript{\emph{a}} Diagrams of the local regions are presented in Figure \ref{fig:boxes};
		\textsuperscript{\emph{b}} For the global region $D_{l}=D$;
		\textsuperscript{\emph{c}} The numbers in brackets are the errors in the last reported decimal place, and errors are determined as one standard error in the mean.  If no bracketed number is given, the statistical error is less than 0.01.}
	\end{table}

When the color current correlation functions are used to calculate $D_{\alpha,l}$ instead of velocity correlations, the error bars are large, but $D_{\alpha,l}$ and $D_{\alpha}$ still agree to within statistical error. This appears to be independent of the box volume compared to the full volume and is correct even when two of the dimensions are short compared to the dimension of the full system, so the number of particles in the regions are far fewer than the total system. Comparing the relative error bars in the results using color current correlation functions (Figures S2 and S3 of the Supplementary Information), it appears that the size of the error bars is related to the number of particles, irrespective of the region shape, which is consistent with the arguments in Section 1.2 of the Supplementary Material. 

\subsection{Inhomogeneous Liquid System}
\begin{figure}[httb!]
    \centering
    \includegraphics[scale=0.4]{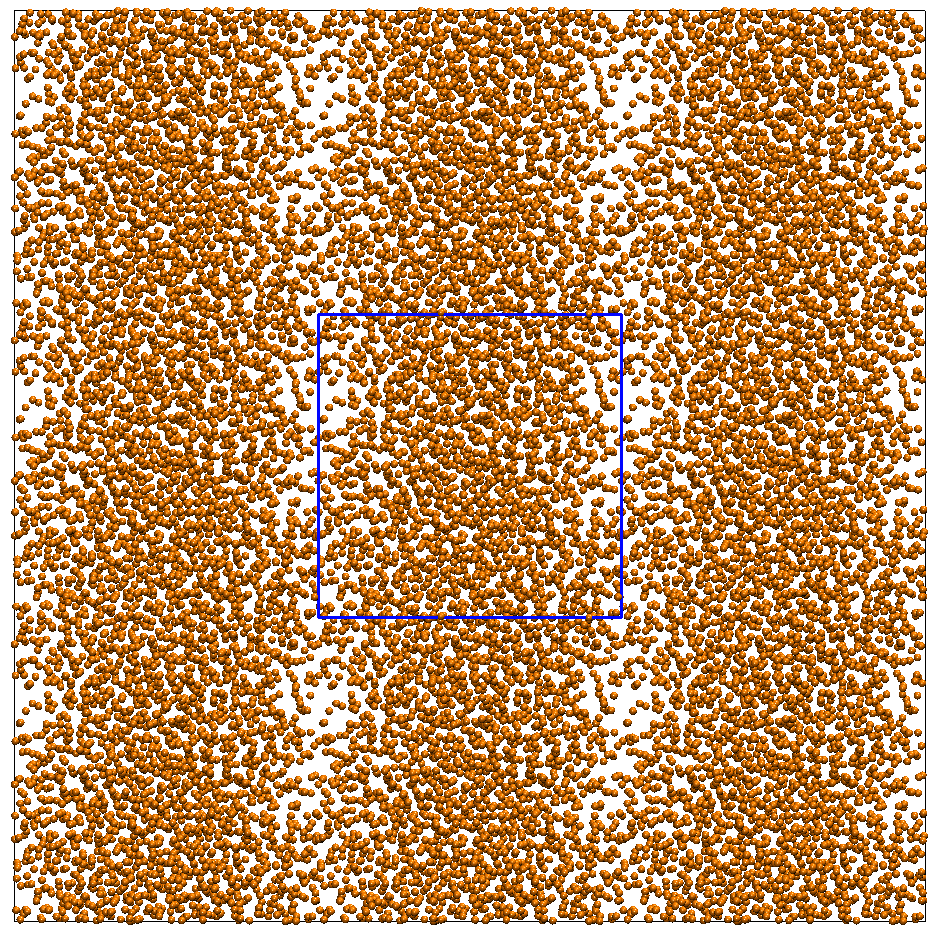}
    \caption{A snapshot of the atoms in the inhomogeneous cubic system. Periodic images in the $x$ and $y$ directions are shown with the blue box showing the unit cell. For this system $N=2048$ and the average density is $\rho = 0.70$. The dimensions of the cubic system were $x = y = z =$ 14.30. The density profile was formed by application of a sinusoidal force in the $x$ direction: $F_{sin} = \text{sin}({2 \pi x}/L_{x})$, where $L_{x}=14.3$. }
    \label{fig:periodicinhomogeneous}
\end{figure}

\begin{figure*}[htbp!]  
	\centering
	\includegraphics[scale=0.45]{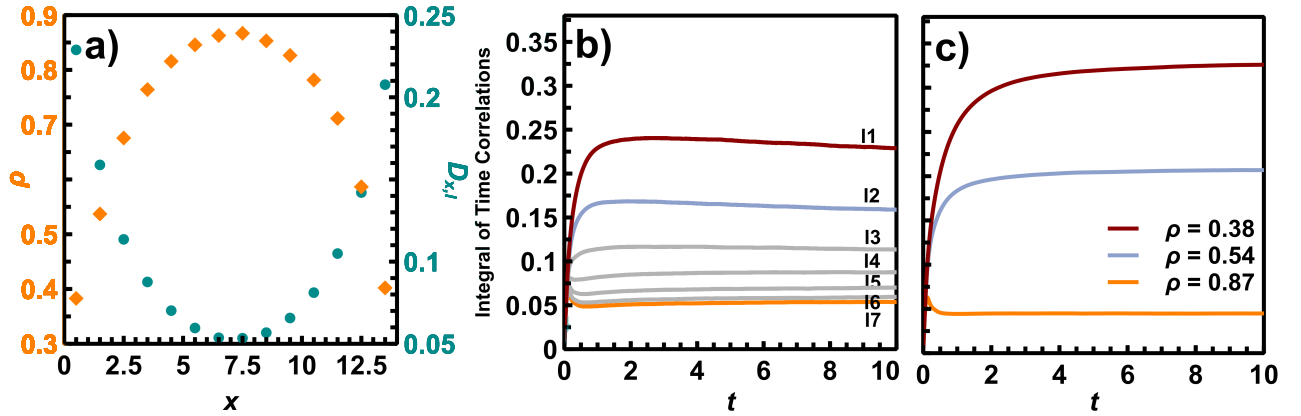}
	\caption{(a) Density profile (orange diamonds) and local diffusion coefficient (teal circles) in the $x$ direction of a liquid system with a sinusoidal force applied $F_{sin} = \text{sin}({2 \pi x}/L_{x})$, where $L_{x}=14.3$ is the length of the $x$ dimension. The local regions that are investigated have dimension $\Delta{x}$ = 1.00, $\Delta y$ = 14.30, $\Delta z$ = 14.30 and are labeled from $l1$ to $l7$ as $x=0...7$. (b) shows the integrals of velocity time correlation functions required for calculations of $D_{x,l}$ in the inhomogeneous system and (c) $D_{x}$ in the homogeneous liquid systems with various densities. The density of each system was chosen to match the average density of a local region: $l1$: $\rho$ = 0.37, $l2$: $\rho$ = 0.54, and $l7$: $\rho$ = 0.87. The curve colors match the corresponding local region in (b).}
	\label{fig:GK_sindensity}
\end{figure*}

A crucial feature of a useful local diffusion coefficient is that its value must vary locally in an inhomogeneous system. To create a single-component inhomogeneous system, a sinusoidal force was applied in the $x$ direction of the cubic simulation cell of length $L_x =14.30$, by adding a term $ \text{sin}(2\pi x/L_x$) to the momentum equation of motion in (\ref{equationsofmotion}), resulting in a non-uniform density profile, shown in Figure \ref{fig:periodicinhomogeneous} and plotted in orange in Figure \ref{fig:GK_sindensity}a.

The local diffusion coefficient decreases consistently with increasing density, as shown in Figure \ref{fig:GK_sindensity}a. For comparison, we simulated three homogeneous liquid systems which had densities similar to the low density region ($l1$: $\rho = 0.37$), medium density region ($l2$: $\rho$=0.54) and the high density region ($l7$: $\rho$ = 0.87) of the inhomogeneous system. The diffusion coefficients in these systems are similar to the local diffusion coefficients in regions 1,2 and 7 in the inhomogeneous system, seen from the integrals of the velocity correlation functions in Figure \ref{fig:GK_sindensity}c, indicating that the calculated local diffusion coefficients are reasonable.

It is important to note that the time correlations in Figures 4b and 4c with corresponding colors and densities are not expected to exactly agree. This is because the values of the time integrals in Figure 4b are from local regions with an average local density shown in Figure 4a, whereas in Figure 4c, they are based on time integrals of homogeneous systems, so the density is the same everywhere in the box. Region l1 (in red) with an local average density of 0.38 will have regions within it that are slightly more dense and less dense than 0.38, and furthermore, these particles will go outside the region in the correlation time, which has a different density. This is different to the homogeneous system with $\rho$ = 0.38, where the density is uniform. Therefore, depending on the correlation time, the values for the diffusion coefficient may be similar, or significantly different comparing a local region of an inhomogeneous system and a homogeneous system of the same density. However, one would expect in a high density region, the gap between these would decrease, because the correlations of the particles will decay before they interact with regions of different density, which we compare by highlighting the orange lines in Figures 4b and c. 

In the inhomogeneous system, the functional form of the integral of the correlation function in $l7$ appears to be that of a liquid, with a peak before plateauing to its limiting behavior, and notably is comparable in both shape and magnitude to the high-density homogeneous system.
In the lower density region $l1$, the functional form indicates that the fluid is gas-like, and has a similar shape to the low-density homogeneous system, but a quantitative match is not observed in this case. This difference can be attributed to the fact that the density gradient is high so the density across $l1$ varies greatly. This demonstrates that the dependence of the diffusion coefficient on the density is not linear and illustrates why accurate methods are required to measure local diffusion coefficients.  

There is an anisotropy in the global diffusion  such that $D_{x}$ is lower than  $D_{y}$ and $D_{z}$, which coincide (shown in Figure S4).  
Comparing the local diffusion coefficients in the $x$ direction with the $y$ and $z$ directions (plotted in Figure S5), we find that the $y$ and the $z$ directions have higher diffusion coefficients, which however are still lower than the homogeneous systems. This can be attributed to the direction of motion relative to the density gradient in the region. For a particle moving in the $x$ direction, the density gradient changes sharply during the correlation time. The density is homogeneous along the $y$ and $z$ directions, so if the particles move along the Cartesian direction for most of the correlation time, their behavior will be more comparable to the homogeneous system. 

\section{Conclusions}
In this research work, we have extended previous studies of the local TTCF to demonstrate that it is possible to define a local diffusion coefficient through a modified Green-Kubo expression. We demonstrate that the results calculated using this expression agree with those obtained for the global diffusion coefficient using the definitions of the local color current and single-particle velocity correlations, within the statistical error. Furthermore, this definition is independent of the shape and size of the local region and the statistical errors are dependent on the number of particles in the region. Unlike previously proposed definitions, the choice of particles to include in the correlation functions is fully prescribed by the theory, and arbitrary choices are not necessary, making the definition valid for any type of inhomogeneity explored and for any arbitrary region, $l$. This means that the user could define, for example a region of a monolayer of a gas or region extending a certain depth from a surface, which makes this method valuable for investigating local self-diffusion coefficients using molecular dynamics simulations.   

We note that in cases where anomalous diffusion occurs, the integrals of the correlation functions will decay to zero (subdiffusive) or not converge (superdiffusive) just as occurs for standard Green-Kubo relations in bulk systems. Similar extensions that are applied for global diffusivity will be required to treat such systems locally.\cite{Dechant2014,Metzler2014,Sahoo2022,Lu2022}. In the systems studied here, this is not an issue. However, for other systems examination of the convergence of the integrals could be used to identify regions of the system where diffusion is anomalous. 

In a proof-of-concept, we show that the local velocity correlations can be used to give local diffusion coefficients that are plausible based on their comparison with homogeneous systems of equivalent density to the local region. Therefore, equilibrium simulations can be used to measure linear transport coefficients in molecular dynamics simulation and can be readily applied to measure nanoscale phenomena in simulations. However, we note that although the diffusion coefficients in local regions are similar to those of a bulk system with the same average density, they are not the same.  The differences reflect the fact that the distribution of velocities in a local region of a highly inhomogeneous system will not be the same as in a bulk system and demonstrates the need for a local expression that properly captures this difference. Conveniently, the new expression for the local diffusion coefficient is easy to implement in the post-processing of simulation data.

\begin{acknowledgement}
	
	The authors thank Dr Emily V. Kahl for her assistance with this project. They thank the Australian Research Council for support of this project through the Discovery program (FL190100080). MAH acknowledges support from the Australian Government through an Australian Government Research Training Program Scholarship. The authors acknowledge access to computational resources at the NCI National Facility through the National Computational Merit Allocation Scheme supported by the Australian Government, and support through resource provided by the Pawsey Supercomputing Centre with funding from the Australian Government and the Government of Western Australia. They also acknowledge support from the Queensland Cyber Infrastructure Foundation (QCIF) and the University of Queensland Research Computing Centre (RCC).
\end{acknowledgement}


\clearpage
\setcounter{section}{0}
\setcounter{figure}{0}
\setcounter{equation}{0}
\usetagform{eqt}
{\centering
\section*{SUPPLEMENTARY INFORMATION}}

\section{Supplementary theory}
\subsection{Equivalence of local color current and velocity autocorrelations in the large system limit}
Consider the local color current,
\begin{eqnarray}
	\label{color current}
	J_{x,l}(t)=\frac{\sum_i^N c_i v_{xi}(t)S_l(\bmat{q}_i(t))}{N_l}
\end{eqnarray}
where $S_l(\bmat{q}_i(t))$ is a switch which is 1 when particle $i$  is in the bin of interest, and 0 when it is not, and $N_l$ is the average number of particles in the bin.  This will give the global color current when there is a single bin, in which case $S_l(\bmat{q}_i(t))=1, \forall i,t$ and $N_l=N$.  The time correlation function of the color current at time 0 and the local color current at time $t$ is then given by,
\begin{eqnarray}
	NN_l\langle J_x(0)J_{x,l}(t)\rangle&=&\left\langle \sum_i^N c_i v_{xi}(0) \sum_j^N c_j v_{xj}(t)S_l({\bmat{q}}_j(t))\right\rangle  \\
	&=& \sum_i^N \sum_j^N c_i c_j \left\langle v_{xi}(0)v_{xj}(t)S_l({\bmat{q}}_j(t))\right\rangle \label{JJ3}\\
	&=& \left\langle v_{x1}(0)v_{x1}(t)S_l({\bmat{q}}_1(t))\right\rangle \sum_i^N c_i^2 \nonumber\\&& +\left\langle v_{x1}(0)v_{x2}(t)S_l({\bmat{q}}_2(t))\right\rangle \sum_i^N   \sum_{j \ne i}^N c_i c_j \label{JJ4}
\end{eqnarray}
where the last equality can be made if all particles are identical except for their color label.

If $c_i=(-1)^i$ and N is even then substituting into (S\ref{JJ4}) and dividing by $N$ gives, for all $i,j$ such that $i\ne j$, 
\begin{eqnarray}
	\label{Jv}
	N_l\langle J_x(0)J_{x,l}(t)\rangle  &=&\left\langle v_{xi}(0)v_{xi}(t)S_l({\bmat{q}}_i(t))\right\rangle - \left\langle v_{xi}(0)v_{xj}(t)S_l({\bmat{q}}_j(t))\right\rangle.
\end{eqnarray}

Now consider the case where $\sum_i^N v_{x,i}=0$ (that is, no streaming velocity).  Then, 
\begin{eqnarray}
	\lefteqn{\left\langle v_{xi}(0)v_{xj}(t)S_l({\bmat{q}}_j(t))\right\rangle
		=-\langle \sum_{k \ne i}v_{xk}(0)v_{xj}(t)S_l({\bmat{q}}_j(t))\rangle}\\
	&=&-\langle v_{xj}(0)v_{xj}(t)S_l({\bmat{q}}_j(t))\rangle-\sum_{k \ne i,j}\langle v_{xk}(0)v_{xj}(t)S_l({\bmat{q}}_j(t))\rangle\\
	&=&-\left\langle v_{xj}(0)v_{xj}(t)S_l({\bmat{q}}_j(t))\right\rangle-(N-2)\left\langle v_{xi}(0)v_{xj}(t)S_l({\bmat{q}}_j(t))\right\rangle.
\end{eqnarray}
Rearranging,
\begin{eqnarray}
	\left\langle v_{xi}(0)v_{xj}(t)S_l({\bmat{q}}_j(t))\right\rangle&=&-\frac{1}{N-1}\left\langle v_{xi}(0)v_{xi}(t)S_l({\bmat{q}}_i(t))\right\rangle.
\end{eqnarray}
Substituting into (S\ref{Jv}),
\begin{eqnarray}
	N_l\langle J_x(0)J_{x,l}(t)\rangle
	=\left(1+\frac{1}{N-1}\right)\left\langle v_{xi}(0)v_{xi}(t)S_l({\bmat{q}}_i(t))\right\rangle, \quad \forall i
\end{eqnarray} 
and in the large $N$ limit,
\begin{eqnarray}
	\label{local_equality}
	N_l\langle J_x(0)J_{x,l}(t)\rangle &=&\left\langle v_{xi}(0)v_{xi}(t)S_l({\bmat{q}}_i(t))\right\rangle, \quad \forall i.
\end{eqnarray}

If we have a case where $\sum_i^N v_{x,i}\ne0$ but $\left\langle\sum_i^N v_{x,i}\right\rangle=0$, we can observe that the velocity of a particle can only be correlated with other particles within the same sound cone \cite{Evans2008b}. In the thermodynamic limit, there will be infinitely more particles outside the sound cone than within it and therefore if the particles are indistinguishable, $\left\langle v_{xi}(0)v_{xj}(t)S_l({\bmat{q}}_j(t))\right\rangle=0$ as $N\rightarrow \infty$, and (S\ref{local_equality}) is obtained from (S\ref{Jv}) directly.  

The local diffusion coefficient is,
\begin{eqnarray}
	\label{diffusion}
	D_{x,l}&=& N\int^\infty_0 ds \langle {J}_{x}(0) {J}_{x,l}(s) \rangle  \\ 
	&=&\frac{N}{N_l}\int^\infty_0 ds \langle {v}_{x,i}(0) {v}_{x,i}(s)S_l({\bmat{q}}_i(s)) \rangle\\
	&=& \int^\infty_0 ds  \langle {v}_{x,i}(0) {v}_{x,i,l}(s) \rangle
\end{eqnarray}
where $ \langle {v}_{x,i}(0) {v}_{x,i,l}(s) \rangle  \equiv N/N_l \langle {v}_{x,i}(0) {v}_{x,i}(s)S_l({\bmat{q}}_i(s)) \rangle $

\subsection{Comment on statistical errors}
Consider the case where there is one bin ($N_l=N$). If (S\ref{JJ3}) is used to evaluate the correlations, a sum over $N(N-1)$ terms $\pm \langle v_{xi}(0)v_{xj}(t)S_l({\bmat{q}}_j(t))\rangle, i\ne j$ is carried out. If the distribution of the values from the $M$ samples has a variance $\sigma_M^2$, the sum of these terms will have variance $N(N-1)\sigma_M^2$, and a standard deviation $\sqrt{N(N-1})\:\sigma_M$. As evident from (S\ref{JJ3}), this is divided by $N_l$ so the contribution to the standard deviation of D will be $\sqrt{(N-1)/N}\:\sigma_M$.  In contrast, using similar arguments, the standard deviation due to only the diagonal terms, $\left\langle v_{xi}(0)v_{xi}(t)S_l({\bmat{q}}_i(t))\right\rangle$,  will shrink as $\sqrt{1/N}\:\sigma_M$.  When $N_l\ne N$, there will be $N_l(N-1)$ terms $\langle v_{xi}(0)v_{xj}(t)S_j(\bmat{q}_j(t))\rangle$ that are not identically zero for finite $M$, and the sum of these terms will have a variance of approximately $N_l(N-1)\sigma_M^2$ and standard deviation of $\sqrt{N_l(N-1)}\:\sigma_M$.  Therefore, using (S\ref{JJ3}), the standard deviation due to this contribution to $D$ will be $\sqrt{(N-1)/N_l}\:\sigma_M$ whereas the contribution from the diagonal terms will shrink as $\sqrt{1/N_l}\:\sigma_M$.  

This makes it very difficult to numerically determine the self diffusion coefficient from the color current time autocorrelation function as the error does not drop as the number of particles does, and even more difficult for local regions where the error increases with the number of particles in the system if $N_l$ remains fixed.  This means that the number of samples needs to increase with the number of particles in the system if the same statistical error is to be obtained.  This is not a problem when the velocity time autocorrelation function is used, with the statistical error in the global and local diffusion coefficients dropping when the number of particles in the system or bin increase, respectively.

\subsection{Equivalence of global-local and local-global correlation functions}
We can write the velocity correlation function in (12) of the main text as a phase space integral
\begin{eqnarray}
	\label{corrfun}
	\langle {v}_{\alpha,i}(0) {v}_{\alpha,i,l}(t) \rangle_{\text{eq}} = \int d\mathcal{S}^t\mathbf{\Gamma} f(\mathcal{S}^t\mathbf{\Gamma})v_{\alpha,i}(\mathbf{\Gamma})v_{\alpha,i}(\mathcal{S}^t\mathbf{\Gamma})S_{i}(\mathcal{S}^t\mathbf{\Gamma})
\end{eqnarray}
where $\mathcal{S}^t$ is the propagator which advances the phase space position forwards in time by a duration $t$ according to the equations of motion (not to be confused with the switch function $S_i$). The phase space distribution function $f$ is not explicitly time dependent because we consider a system at equilibrium. The equilibrium distribution is time reversal symmetric, so $f(\mathcal{S}^t\mathbf{\Gamma})=f(M^T\mathcal{S}^t\mathbf{\Gamma})$,
where $M^T$ is the time reversal map, $M^T(\mathbf{q},\mathbf{p)} = (\mathbf{q},-\mathbf{p)}$. Velocities are reversed under time reversal, so ${v}_{\alpha,i}(M^{T}\mathcal{S}^t\mathbf{\Gamma})=-{v}_{\alpha,i}(\mathcal{S}^t\mathbf{\Gamma})$. The switch function depends only on position, so $S_{i}(M^{T}\mathcal{S}^t\mathbf{\Gamma})=S_{i}(\mathcal{S}^t\mathbf{\Gamma})$. The equations of motion are time reversible, $M^{T}\mathcal{S}^tM^{T}\mathcal{S}^t\mathbf{\Gamma}=\mathbf{\Gamma}$,
so ${v}_{\alpha,i}(M^{T}\mathcal{S}^tM^{T}\mathcal{S}^{t_{1}}\mathbf{\Gamma})={v}_{\alpha,i}(\mathbf{\Gamma})$.
Therefore, ${v}_{\alpha,i}(\mathcal{S}^tM^{T}\mathcal{S}^t\mathbf{\Gamma})=-{v}_{\alpha,i}(\mathbf{\Gamma})$.
Then substituting into (S\ref{corrfun}),
\begin{eqnarray}
	\label{corrfun2}
	\langle {v}_{\alpha,i}(0) {v}_{\alpha,i,l}(t) \rangle_{\text{eq}} = \int d\mathcal{S}^t\mathbf{\Gamma} f(M^T\mathcal{S}^t\mathbf{\Gamma})v_{\alpha,i}(\mathcal{S}^tM^TS^t\mathbf{\Gamma})v_{\alpha,i}(M^T\mathcal{S}^t\mathbf{\Gamma})S_{i}(M^T\mathcal{S}^t\mathbf{\Gamma}).
\end{eqnarray}
Let $\mathbf{\Gamma}^{*}=M^{T}\mathcal{S}^t \mathbf{\Gamma}$. Since the Jacobian of the time reversal map is one, $d\mathcal{S}^t\mathbf{\Gamma} = dM^{T}\mathbf{\Gamma}^{*}=d\mathbf{\Gamma}^{*}$. Now, 
\begin{eqnarray}
	\label{corrfun3}
	\langle {v}_{\alpha,i}(0) {v}_{\alpha,i,l}(t) \rangle_{\text{eq}} &=& \int d\mathbf{\Gamma}^* f(\mathbf{\Gamma}^*)v_{\alpha,i}(\mathcal{S}^t\mathbf{\Gamma}^*)v_{\alpha,i}(\mathbf{\Gamma}^*)S_{i}(\mathbf{\Gamma}^*)\\
	&=& \langle {v}_{\alpha,i,l}(0) {v}_{\alpha,i}(t)\rangle_{\text{eq}}.
\end{eqnarray}
Since $t$ is arbitrary, this is true for all times in the integral in (12) of the main text. 
\clearpage
\section{Supplementary Figures}
\renewcommand{\thefigure}{S\arabic{figure}}

\begin{figure}[http!]
	\centering
	\includegraphics[scale=0.6]{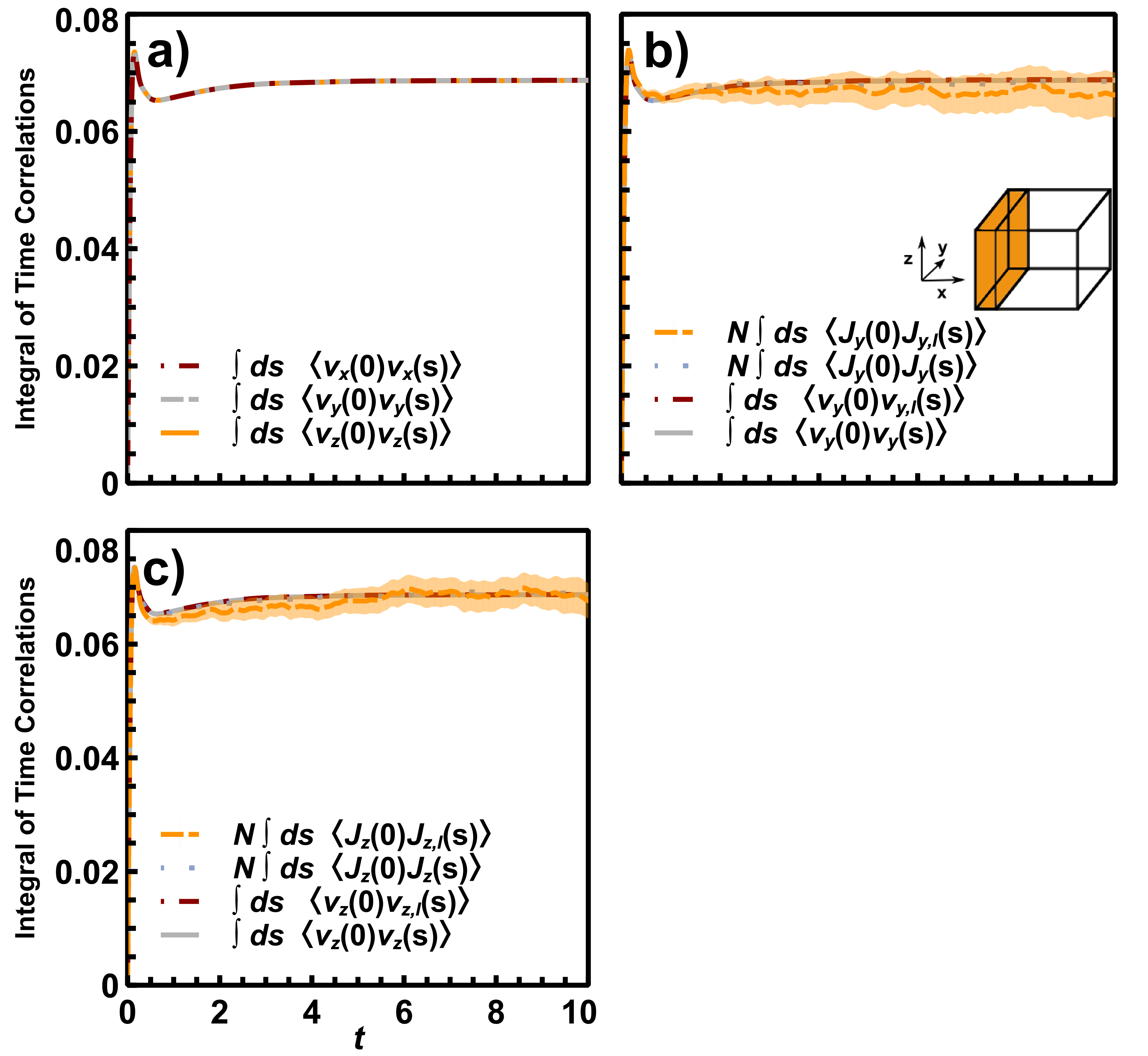}
	\caption{Integrals of the time correlation functions which converge to give the components of the diffusion coefficients at long times in the homogeneous cubic system presented in Figure S1. (a) Results obtained using the global velocity autocorrelations. Note that integrals used to determine $D_{x}$, $D_{y}$ and $D_{z}$ overlap, as expected for the cubic system. (b)-(c) Results obtained using the global and local-global correlations in the color current and the velocity with equations (8) and (12). The asymptotic results give the diffusion coefficients for (b) the $y$ direction and (c) the $z$ direction.  The error bars in light orange show the value of 1 standard error in the mean of the result obtained from correlations of the local color current. Note that the true value of $D$ is the long time limit of each integral, but these simulations are run for finite time. It can be seen that the results obtained using the color current are noisier, however approach the same value of $D_{y}$ and $D_{z}$, respectively. }
	\label{fig:GKTTCFycube}
\end{figure}

\begin{figure}[http!]
	\centering
	\includegraphics[scale=0.6]{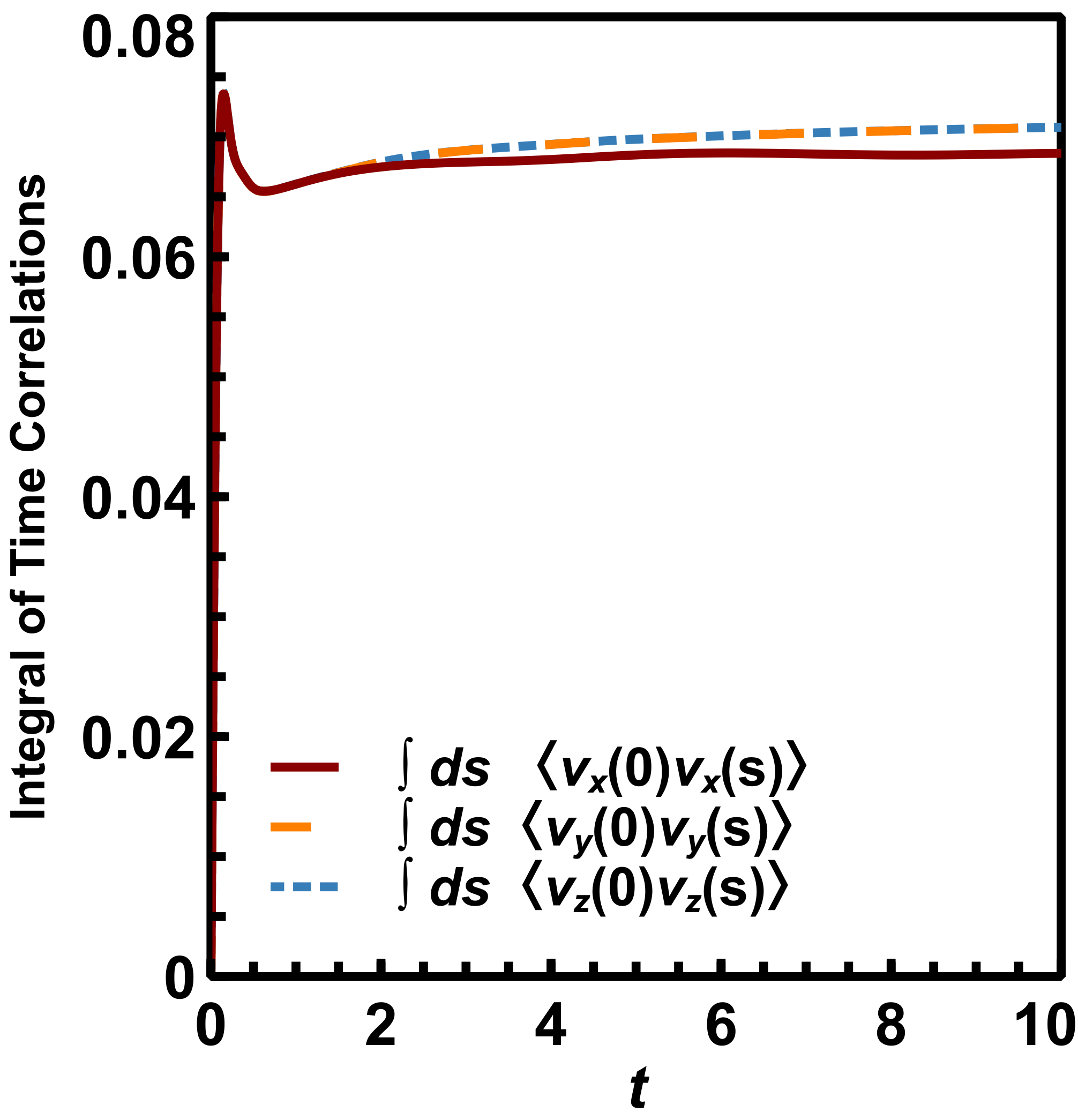}
	\caption{Integrals of the time correlation functions which converge to give the $x$, $y$ and $z$-components of the diffusion coefficients at long times for the non-cubic homogeneous system presented in Figure S1. The results were obtained using the global velocity autocorrelations. It can be seen that the predicted value of $D_{x}$ is lower than $D_{y}$ which is equal to $D_{z}$, and the curves of the data in the $y$ and $z$ directions overlap. The anisotropy of the diffusion is due to the different system size effects in the $x$ direction than in the $y$ and $z$ directions in this non-cubic box. }
	\label{fig:global_GK_anisotropic}
\end{figure}

\begin{figure*}[htbp!]  
	\centering
	\includegraphics[scale=0.45]{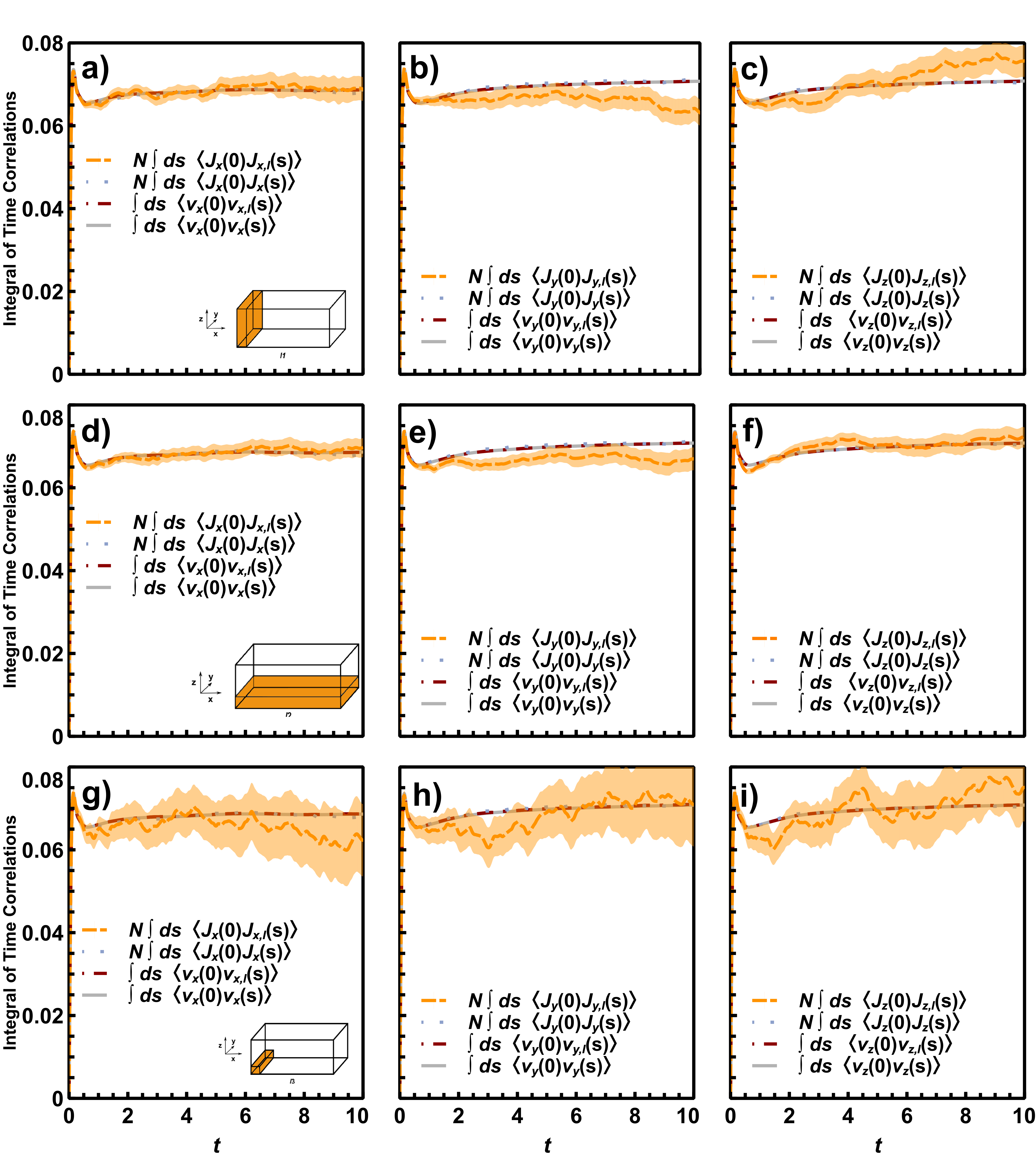}
	\caption{Integrals of the time correlation functions for the calculation of the global and local diffusion coefficients in local regions $l1$ (a) - (c), $l2$ (d)-(f), and $l3$ (g)-(i) for each direction in an liquid system with anisotropic diffusion. Details of the local regions are given in the caption to Figure S1. Note that $D_{\alpha}$ and $D_{\alpha,l}$ are given by the long time limit of each integral. There is a small difference in calculated value of $D_{\alpha}$ between the $x$ and $z$ directions which is due to the simulation box being non-cubic. The standard error in the mean for the local color current is in light orange and is larger for $l3$ due to the smaller region size.}
	\label{fig:GKTTCFan}
\end{figure*}

\begin{figure}[http!]  
	\centering
	\includegraphics[scale=0.45]{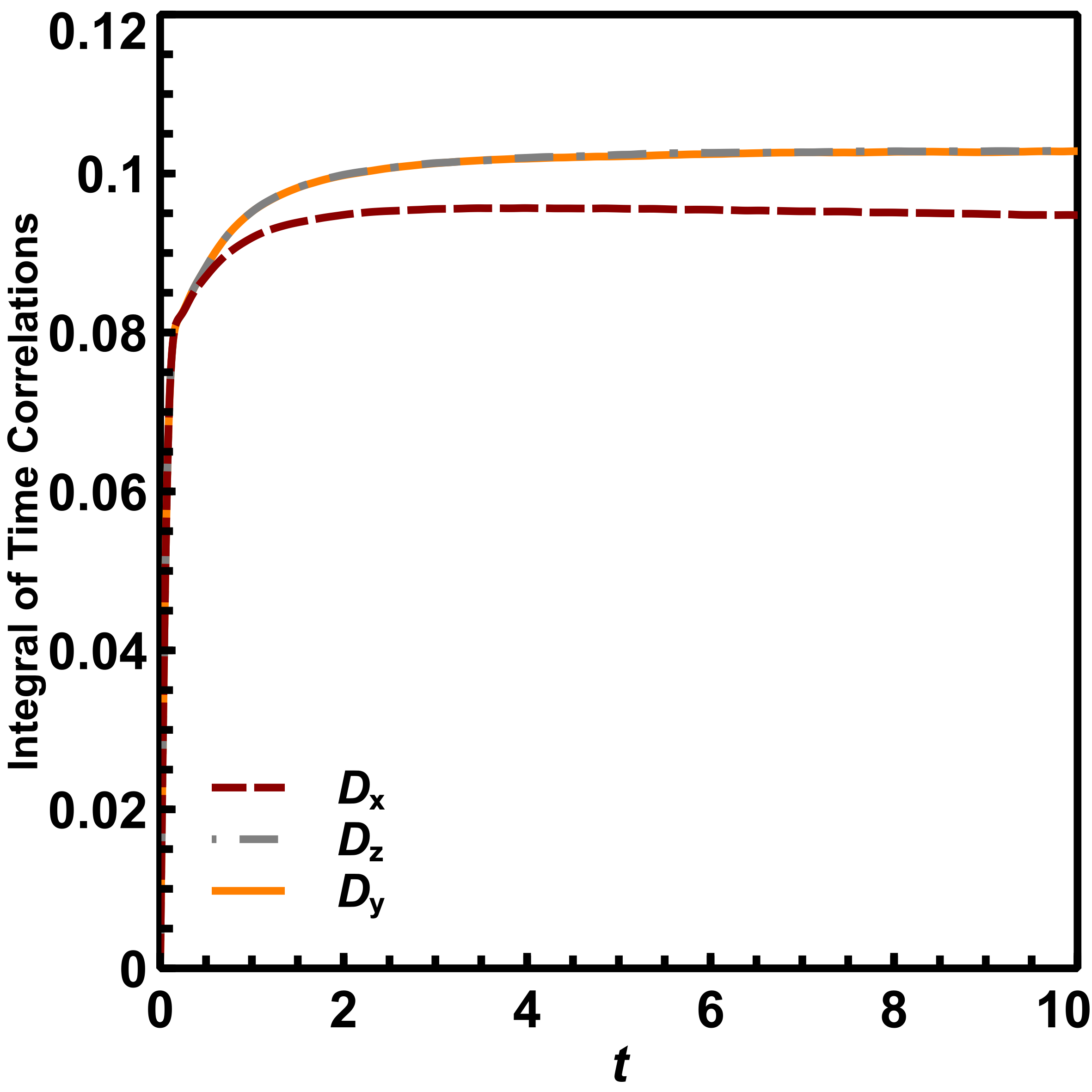}
	\caption{Global velocity autocorrelations in each direction in an inhomogeneous liquid system. It can be seen that due to the density profile created by the force, the global diffusion is not isotropic with $D_x < D_{y} = D_{z}$ (the curves for $D_y$ and $D_z$ overlap).}
	\label{fig:globalGK_inhomogeneous}
\end{figure}

\begin{figure}[http!]  
	\centering
	\includegraphics[scale=0.45]{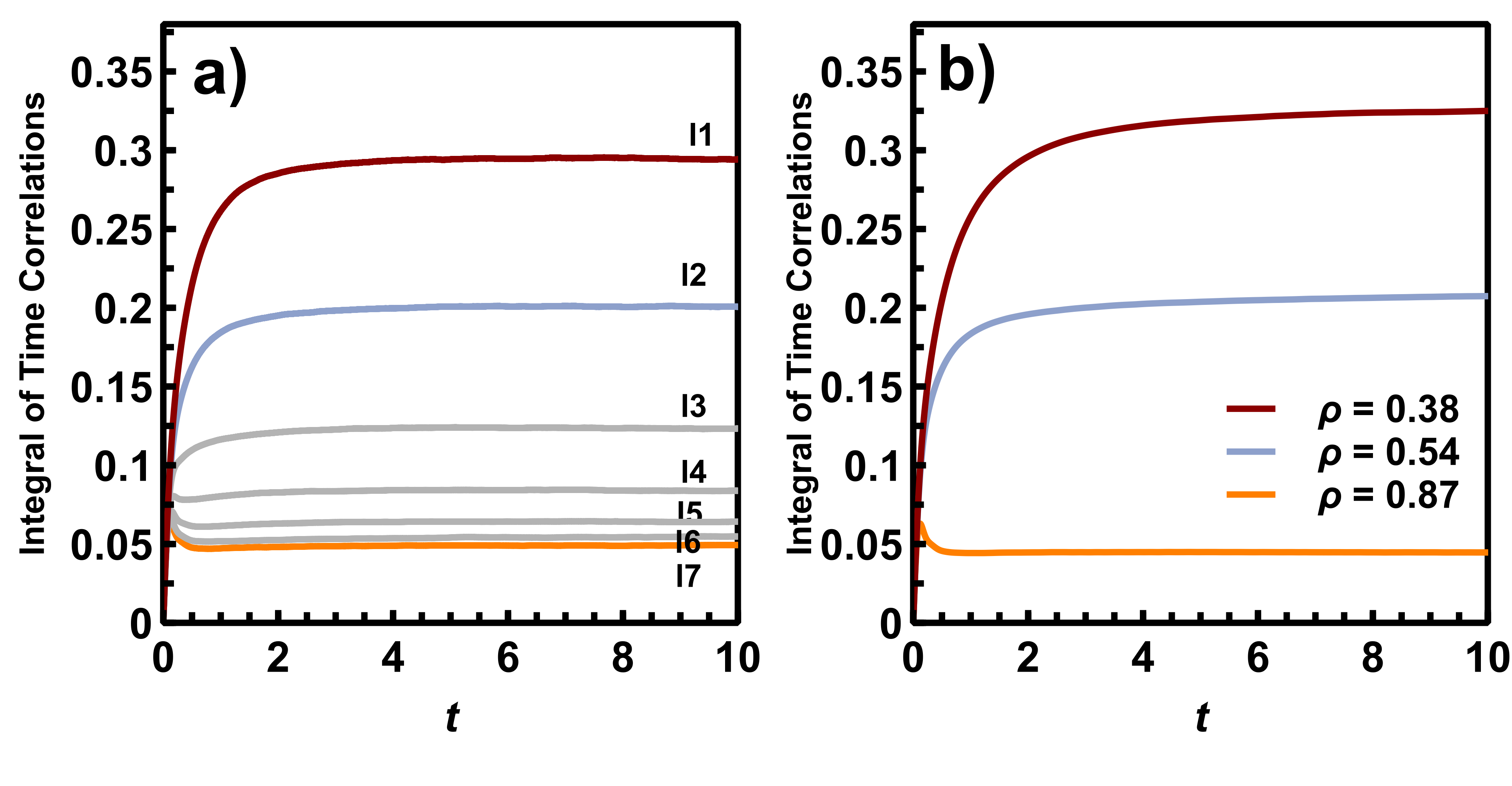}
	\caption{Integrals of time correlation functions required for calculations of $D_y$ in (a) an inhomogeneous liquid system with an average density $\rho$ = 0.70 and (b) homogeneous liquid system at three densities. (a) shows the integrals of local-global velocity correlations for bins of dimensions $\Delta{x}$ = 1, $\Delta y$ = 14.30, $\Delta z$ = 14.30 in the $y$ and $z$ directions respectively. The bins are labeled from $l1$ to $l7$ as $x=0...7$ with $F_{sin} = \text{sin}({2 \pi x}/L_{x})$. (b) shows the calculated $D_{y}$ of a pure system with a bulk density equal to the average local density in $l1$ ($\rho$=0.37, maroon), $l2$ ($\rho$ = 0.54, blue) and $l7$ ($\rho$ = 0.87, orange) calculated using the global correlations in the velocity. The results for $D_z$ appear identical and are not shown here.}
	\label{fig:GK_siny}
\end{figure}
\bibliography{references}

\providecommand{\latin}[1]{#1}
\makeatletter
\providecommand{\doi}
  {\begingroup\let\do\@makeother\dospecials
  \catcode`\{=1 \catcode`\}=2 \doi@aux}
\providecommand{\doi@aux}[1]{\endgroup\texttt{#1}}
\makeatother
\providecommand*\mcitethebibliography{\thebibliography}
\csname @ifundefined\endcsname{endmcitethebibliography}
  {\let\endmcitethebibliography\endthebibliography}{}
\begin{mcitethebibliography}{35}
\providecommand*\natexlab[1]{#1}
\providecommand*\mciteSetBstSublistMode[1]{}
\providecommand*\mciteSetBstMaxWidthForm[2]{}
\providecommand*\mciteBstWouldAddEndPuncttrue
  {\def\EndOfBibitem{\unskip.}}
\providecommand*\mciteBstWouldAddEndPunctfalse
  {\let\EndOfBibitem\relax}
\providecommand*\mciteSetBstMidEndSepPunct[3]{}
\providecommand*\mciteSetBstSublistLabelBeginEnd[3]{}
\providecommand*\EndOfBibitem{}
\mciteSetBstSublistMode{f}
\mciteSetBstMaxWidthForm{subitem}{(\alph{mcitesubitemcount})}
\mciteSetBstSublistLabelBeginEnd
  {\mcitemaxwidthsubitemform\space}
  {\relax}
  {\relax}

\bibitem[Todd and Evans(1995)Todd, and Evans]{Todd1995}
Todd,~B.~D.; Evans,~D.~J. The heat flux vector for highly inhomogeneous
  nonequilibrium fluids in very narrow pores. \emph{The Journal of Chemical
  Physics} \textbf{1995}, \emph{103}, 9804--9809\relax
\mciteBstWouldAddEndPuncttrue
\mciteSetBstMidEndSepPunct{\mcitedefaultmidpunct}
{\mcitedefaultendpunct}{\mcitedefaultseppunct}\relax
\EndOfBibitem
\bibitem[Hansen \latin{et~al.}(2007)Hansen, Daivis, Travis, and
  Todd]{Hansen2007}
Hansen,~J.~S.; Daivis,~P.~J.; Travis,~K.~P.; Todd,~B.~D. Parameterization of
  the nonlocal viscosity kernel for an atomic fluid. \emph{Physical Review E}
  \textbf{2007}, \emph{76}, 041121\relax
\mciteBstWouldAddEndPuncttrue
\mciteSetBstMidEndSepPunct{\mcitedefaultmidpunct}
{\mcitedefaultendpunct}{\mcitedefaultseppunct}\relax
\EndOfBibitem
\bibitem[Hoang and Galliero(2012)Hoang, and Galliero]{Galliero2012}
Hoang,~H.; Galliero,~G. Local viscosity of a fluid confined in a narrow pore.
  \emph{Phys. Rev. E} \textbf{2012}, \emph{86}, 021202\relax
\mciteBstWouldAddEndPuncttrue
\mciteSetBstMidEndSepPunct{\mcitedefaultmidpunct}
{\mcitedefaultendpunct}{\mcitedefaultseppunct}\relax
\EndOfBibitem
\bibitem[Hoang and Galliero(2012)Hoang, and Galliero]{Hoang2012}
Hoang,~H.; Galliero,~G. Shear viscosity of inhomogeneous fluids. \emph{The
  Journal of Chemical Physics} \textbf{2012}, \emph{136}, 124902\relax
\mciteBstWouldAddEndPuncttrue
\mciteSetBstMidEndSepPunct{\mcitedefaultmidpunct}
{\mcitedefaultendpunct}{\mcitedefaultseppunct}\relax
\EndOfBibitem
\bibitem[Buhn \latin{et~al.}(2004)Buhn, Bopp, and Hampe]{Buhn2004}
Buhn,~J.~B.; Bopp,~P.~A.; Hampe,~M.~J. {A molecular dynamics study of a
  liquid-liquid interface: Structure and dynamics}. \emph{Fluid Phase
  Equilibria} \textbf{2004}, \emph{224}, 221--230\relax
\mciteBstWouldAddEndPuncttrue
\mciteSetBstMidEndSepPunct{\mcitedefaultmidpunct}
{\mcitedefaultendpunct}{\mcitedefaultseppunct}\relax
\EndOfBibitem
\bibitem[Buhn \latin{et~al.}(2006)Buhn, Bopp, and Hampe]{Buhn2006a}
Buhn,~J.~B.; Bopp,~P.~A.; Hampe,~M.~J. {Structural and dynamical properties of
  liquid-liquid interfaces: A systematic molecular dynamics study}.
  \emph{Journal of Molecular Liquids} \textbf{2006}, \emph{125}, 187--196\relax
\mciteBstWouldAddEndPuncttrue
\mciteSetBstMidEndSepPunct{\mcitedefaultmidpunct}
{\mcitedefaultendpunct}{\mcitedefaultseppunct}\relax
\EndOfBibitem
\bibitem[Braga \latin{et~al.}(2014)Braga, Galindo, and M{\"{u}}ller]{Braga2014}
Braga,~C.; Galindo,~A.; M{\"{u}}ller,~E.~A. {Nonequilibrium molecular dynamics
  simulation of diffusion at the liquid-liquid interface}. \emph{Journal of
  Chemical Physics} \textbf{2014}, \emph{141}, 154101\relax
\mciteBstWouldAddEndPuncttrue
\mciteSetBstMidEndSepPunct{\mcitedefaultmidpunct}
{\mcitedefaultendpunct}{\mcitedefaultseppunct}\relax
\EndOfBibitem
\bibitem[Chilukoti \latin{et~al.}(2016)Chilukoti, Kikugawa, and
  Ohara]{Chilukoti2016}
Chilukoti,~H.~K.; Kikugawa,~G.; Ohara,~T. {Structure and mass transport
  characteristics at the intrinsic liquid-vapor interfaces of alkanes}.
  \emph{Journal of Physical Chemistry B} \textbf{2016}, \emph{120},
  7207--7216\relax
\mciteBstWouldAddEndPuncttrue
\mciteSetBstMidEndSepPunct{\mcitedefaultmidpunct}
{\mcitedefaultendpunct}{\mcitedefaultseppunct}\relax
\EndOfBibitem
\bibitem[Buhn \latin{et~al.}(2004)Buhn, Bopp, and Hampe]{Buhn2004a}
Buhn,~J.~B.; Bopp,~P.~A.; Hampe,~M.~J. {A molecular dynamics study of a
  liquid-liquid interface: Structure and dynamics}. \emph{Fluid Phase
  Equilibria} \textbf{2004}, \emph{224}, 221--230\relax
\mciteBstWouldAddEndPuncttrue
\mciteSetBstMidEndSepPunct{\mcitedefaultmidpunct}
{\mcitedefaultendpunct}{\mcitedefaultseppunct}\relax
\EndOfBibitem
\bibitem[Liu \latin{et~al.}(2004)Liu, Harder, and Berne]{Liu2004}
Liu,~P.; Harder,~E.; Berne,~B.~J. {On the calculation of diffusion coefficients
  in confined fluids and interfaces with an application to the liquid-vapor
  interface of water}. \emph{Journal of Physical Chemistry B} \textbf{2004},
  \emph{108}, 6595--6602\relax
\mciteBstWouldAddEndPuncttrue
\mciteSetBstMidEndSepPunct{\mcitedefaultmidpunct}
{\mcitedefaultendpunct}{\mcitedefaultseppunct}\relax
\EndOfBibitem
\bibitem[Chilukoti \latin{et~al.}(2015)Chilukoti, Kikugawa, and
  Ohara]{Chilukoti2015}
Chilukoti,~H.~K.; Kikugawa,~G.; Ohara,~T. {Self-diffusion coefficient and
  structure of binary n-alkane mixtures at the liquid-vapor interfaces}.
  \emph{Journal of Physical Chemistry B} \textbf{2015}, \emph{119},
  13177--13184\relax
\mciteBstWouldAddEndPuncttrue
\mciteSetBstMidEndSepPunct{\mcitedefaultmidpunct}
{\mcitedefaultendpunct}{\mcitedefaultseppunct}\relax
\EndOfBibitem
\bibitem[Colmenares \latin{et~al.}(2009)Colmenares, L{\'{o}}pez, and
  Olivares-Rivas]{Colmenares2009}
Colmenares,~P.~J.; L{\'{o}}pez,~F.; Olivares-Rivas,~W. {Molecular dynamics and
  analytical Langevin equation approach for the self-diffusion constant of an
  anisotropic fluid}. \emph{Physical Review E - Statistical, Nonlinear, and
  Soft Matter Physics} \textbf{2009}, \emph{80}, 1--9\relax
\mciteBstWouldAddEndPuncttrue
\mciteSetBstMidEndSepPunct{\mcitedefaultmidpunct}
{\mcitedefaultendpunct}{\mcitedefaultseppunct}\relax
\EndOfBibitem
\bibitem[{Mercier Franco} \latin{et~al.}(2016){Mercier Franco}, Castier, and
  Economou]{MercierFranco2016}
{Mercier Franco},~L.~F.; Castier,~M.; Economou,~I.~G. {Diffusion in homogeneous
  and in inhomogeneous media: A new unified approach}. \emph{Journal of
  Chemical Theory and Computation} \textbf{2016}, \emph{12}, 5247--5255\relax
\mciteBstWouldAddEndPuncttrue
\mciteSetBstMidEndSepPunct{\mcitedefaultmidpunct}
{\mcitedefaultendpunct}{\mcitedefaultseppunct}\relax
\EndOfBibitem
\bibitem[Benjamin(1992)]{Benjamin1992}
Benjamin,~I. {Theoretical study of the water/1,2-dichloroethane interface:
  Structure, dynamics, and conformational equilibria at the liquid-liquid
  interface}. \emph{The Journal of Chemical Physics} \textbf{1992}, \emph{97},
  1432--1445\relax
\mciteBstWouldAddEndPuncttrue
\mciteSetBstMidEndSepPunct{\mcitedefaultmidpunct}
{\mcitedefaultendpunct}{\mcitedefaultseppunct}\relax
\EndOfBibitem
\bibitem[Chio and Tse(2020)Chio, and Tse]{Chio2020}
Chio,~C.~C.; Tse,~Y. L.~S. {Hindered diffusion near fluid-solid interfaces:
  Comparison of molecular dynamics to continuum hydrodynamics}. \emph{Langmuir}
  \textbf{2020}, \emph{36}, 9412--9423\relax
\mciteBstWouldAddEndPuncttrue
\mciteSetBstMidEndSepPunct{\mcitedefaultmidpunct}
{\mcitedefaultendpunct}{\mcitedefaultseppunct}\relax
\EndOfBibitem
\bibitem[F{\'{a}}bi{\'{a}}n \latin{et~al.}(2016)F{\'{a}}bi{\'{a}}n,
  Sen{\'{c}}anski, Cvijeti{\'{c}}, Jedlovszky, and Horvai]{Fabian2016}
F{\'{a}}bi{\'{a}}n,~B.; Sen{\'{c}}anski,~M.~V.; Cvijeti{\'{c}},~I.~N.;
  Jedlovszky,~P.; Horvai,~G. {Dynamics of the water molecules at the intrinsic
  liquid surface as seen from molecular dynamics simulation and identification
  of truly interfacial molecules analysis}. \emph{Journal of Physical Chemistry
  C} \textbf{2016}, \emph{120}, 8578--8588\relax
\mciteBstWouldAddEndPuncttrue
\mciteSetBstMidEndSepPunct{\mcitedefaultmidpunct}
{\mcitedefaultendpunct}{\mcitedefaultseppunct}\relax
\EndOfBibitem
\bibitem[F{\'{a}}bi{\'{a}}n \latin{et~al.}(2020)F{\'{a}}bi{\'{a}}n, Horvai,
  Sega, and Jedlovszky]{Fabian2020}
F{\'{a}}bi{\'{a}}n,~B.; Horvai,~G.; Sega,~M.; Jedlovszky,~P. {Single Particle
  Dynamics at the Liquid-Liquid Interface. Molecular Dynamics Simulation Study
  of the Water-CCl4 System}. \emph{Journal of Physical Chemistry C}
  \textbf{2020}, \emph{124}, 2039--2049\relax
\mciteBstWouldAddEndPuncttrue
\mciteSetBstMidEndSepPunct{\mcitedefaultmidpunct}
{\mcitedefaultendpunct}{\mcitedefaultseppunct}\relax
\EndOfBibitem
\bibitem[Zaragoza \latin{et~al.}(2019)Zaragoza, Gonzalez, Joly,
  L{\'{o}}pez-Montero, Canales, Benavides, and Valeriani]{Zaragoza2019}
Zaragoza,~A.; Gonzalez,~M.~A.; Joly,~L.; L{\'{o}}pez-Montero,~I.;
  Canales,~M.~A.; Benavides,~A.~L.; Valeriani,~C. {Molecular dynamics study of
  nanoconfined TIP4P/2005 water: How confinement and temperature affect
  diffusion and viscosity}. \emph{Physical Chemistry Chemical Physics}
  \textbf{2019}, \emph{21}, 13653--13667\relax
\mciteBstWouldAddEndPuncttrue
\mciteSetBstMidEndSepPunct{\mcitedefaultmidpunct}
{\mcitedefaultendpunct}{\mcitedefaultseppunct}\relax
\EndOfBibitem
\bibitem[Olivares-Rivas \latin{et~al.}(2013)Olivares-Rivas, Colmenares, and
  L{\'{o}}pez]{Olivares-Rivas2013}
Olivares-Rivas,~W.; Colmenares,~P.~J.; L{\'{o}}pez,~F. {Direct evaluation of
  the position dependent diffusion coefficient and persistence time from the
  equilibrium density profile in anisotropic fluids}. \emph{Journal of Chemical
  Physics} \textbf{2013}, \emph{139}\relax
\mciteBstWouldAddEndPuncttrue
\mciteSetBstMidEndSepPunct{\mcitedefaultmidpunct}
{\mcitedefaultendpunct}{\mcitedefaultseppunct}\relax
\EndOfBibitem
\bibitem[Vermorel \latin{et~al.}(2017)Vermorel, Oulebsir, and
  Galliero]{Vermorel2017}
Vermorel,~R.; Oulebsir,~F.; Galliero,~G. {Communication: A method to compute
  the transport coefficient of pure fluids diffusing through planar interfaces
  from equilibrium molecular dynamics simulations}. \emph{Journal of Chemical
  Physics} \textbf{2017}, \emph{147}, 1--6\relax
\mciteBstWouldAddEndPuncttrue
\mciteSetBstMidEndSepPunct{\mcitedefaultmidpunct}
{\mcitedefaultendpunct}{\mcitedefaultseppunct}\relax
\EndOfBibitem
\bibitem[Evans \latin{et~al.}(2008)Evans, Searles, and Williams]{Evans2008}
Evans,~D.~J.; Searles,~D.~J.; Williams,~S.~R. {On the fluctuation theorem for
  the dissipation function and its connection with response theory}. \emph{The
  Journal of Chemical Physics} \textbf{2008}, \emph{128}, 14504\relax
\mciteBstWouldAddEndPuncttrue
\mciteSetBstMidEndSepPunct{\mcitedefaultmidpunct}
{\mcitedefaultendpunct}{\mcitedefaultseppunct}\relax
\EndOfBibitem
\bibitem[Evans and Morriss(2008)Evans, and Morriss]{Evans2008b}
Evans,~D.~J.; Morriss,~G.~P. \emph{Statistical mechanics of nonequilibrium
  liquids}, 2nd ed.; Cambridge University Press: Cambridge, 2008\relax
\mciteBstWouldAddEndPuncttrue
\mciteSetBstMidEndSepPunct{\mcitedefaultmidpunct}
{\mcitedefaultendpunct}{\mcitedefaultseppunct}\relax
\EndOfBibitem
\bibitem[Sevick \latin{et~al.}(2008)Sevick, Prabhakar, Williams, and
  Searles]{Sevick2008a}
Sevick,~E.~M.; Prabhakar,~R.; Williams,~S.~R.; Searles,~D.~J. {Fluctuation
  theorems}. \emph{Annual Review of Physical Chemistry} \textbf{2008},
  \emph{59}, 603--633\relax
\mciteBstWouldAddEndPuncttrue
\mciteSetBstMidEndSepPunct{\mcitedefaultmidpunct}
{\mcitedefaultendpunct}{\mcitedefaultseppunct}\relax
\EndOfBibitem
\bibitem[Searles and Evans(2000)Searles, and Evans]{Searles2000b}
Searles,~D.~J.; Evans,~D.~J. Ensemble dependence of the transient fluctuation
  theorem. \emph{The Journal of Chemical Physics} \textbf{2000}, \emph{113},
  3503--3509\relax
\mciteBstWouldAddEndPuncttrue
\mciteSetBstMidEndSepPunct{\mcitedefaultmidpunct}
{\mcitedefaultendpunct}{\mcitedefaultseppunct}\relax
\EndOfBibitem
\bibitem[Evans and Searles(1994)Evans, and Searles]{Evans1994a}
Evans,~D.~J.; Searles,~D.~J. {Equilibrium microstates which generate second law
  violating steady states}. \emph{Physical Review E} \textbf{1994}, \emph{50},
  1645--1648\relax
\mciteBstWouldAddEndPuncttrue
\mciteSetBstMidEndSepPunct{\mcitedefaultmidpunct}
{\mcitedefaultendpunct}{\mcitedefaultseppunct}\relax
\EndOfBibitem
\bibitem[Talaei \latin{et~al.}(2012)Talaei, Reid, and Searles]{Talaei2012}
Talaei,~Z.; Reid,~J.~C.; Searles,~D.~J. {A local dissipation theorem}.
  \emph{Journal of Chemical Physics} \textbf{2012}, \emph{137}, 214110\relax
\mciteBstWouldAddEndPuncttrue
\mciteSetBstMidEndSepPunct{\mcitedefaultmidpunct}
{\mcitedefaultendpunct}{\mcitedefaultseppunct}\relax
\EndOfBibitem
\bibitem[Brookes(2016)]{Brookes2016}
Brookes,~S. Ph.D.\ thesis, Griffith University, 2016\relax
\mciteBstWouldAddEndPuncttrue
\mciteSetBstMidEndSepPunct{\mcitedefaultmidpunct}
{\mcitedefaultendpunct}{\mcitedefaultseppunct}\relax
\EndOfBibitem
\bibitem[Yeh and Hummer(2004)Yeh, and Hummer]{Yeh2004}
Yeh,~I.~C.; Hummer,~G. {System-size dependence of diffusion coefficients and
  viscosities from molecular dynamics simulations with periodic boundary
  conditions}. \emph{Journal of Physical Chemistry B} \textbf{2004},
  \emph{108}, 15873--15879\relax
\mciteBstWouldAddEndPuncttrue
\mciteSetBstMidEndSepPunct{\mcitedefaultmidpunct}
{\mcitedefaultendpunct}{\mcitedefaultseppunct}\relax
\EndOfBibitem
\bibitem[Celebi \latin{et~al.}(2021)Celebi, Jamali, Bardow, Vlugt, and
  Moultos]{Celebi2020}
Celebi,~A.~T.; Jamali,~S.~H.; Bardow,~A.; Vlugt,~T. J.~H.; Moultos,~O.~A.
  Finite-size effects of diffusion coefficients computed from molecular
  dynamics: A review of what we have learned so far. \emph{Molecular
  Simulation} \textbf{2021}, \emph{47}, 831--845\relax
\mciteBstWouldAddEndPuncttrue
\mciteSetBstMidEndSepPunct{\mcitedefaultmidpunct}
{\mcitedefaultendpunct}{\mcitedefaultseppunct}\relax
\EndOfBibitem
\bibitem[Jamali \latin{et~al.}(2020)Jamali, Bardow, Vlugt, and
  Moultos]{Jamali2020}
Jamali,~S.~H.; Bardow,~A.; Vlugt,~T.~J.; Moultos,~O.~A. {Generalized Form for
  Finite-Size Corrections in Mutual Diffusion Coefficients of Multicomponent
  Mixtures Obtained from Equilibrium Molecular Dynamics Simulation}.
  \emph{Journal of Chemical Theory and Computation} \textbf{2020}, \emph{16},
  3799--3806\relax
\mciteBstWouldAddEndPuncttrue
\mciteSetBstMidEndSepPunct{\mcitedefaultmidpunct}
{\mcitedefaultendpunct}{\mcitedefaultseppunct}\relax
\EndOfBibitem
\bibitem[Dechant \latin{et~al.}(2014)Dechant, Lutz, Kessler, and
  Barkai]{Dechant2014}
Dechant,~A.; Lutz,~E.; Kessler,~D.~A.; Barkai,~E. Scaling Green-Kubo Relation
  and Application to Three Aging Systems. \emph{Phys. Rev. X} \textbf{2014},
  \emph{4}, 011022\relax
\mciteBstWouldAddEndPuncttrue
\mciteSetBstMidEndSepPunct{\mcitedefaultmidpunct}
{\mcitedefaultendpunct}{\mcitedefaultseppunct}\relax
\EndOfBibitem
\bibitem[Metzler \latin{et~al.}(2014)Metzler, Jeon, Cherstvy, and
  Barkai]{Metzler2014}
Metzler,~R.; Jeon,~J.-H.; Cherstvy,~A.~G.; Barkai,~E. Anomalous diffusion
  models and their properties: non-stationarity{,} non-ergodicity{,} and ageing
  at the centenary of single particle tracking. \emph{Phys. Chem. Chem. Phys.}
  \textbf{2014}, \emph{16}, 24128--24164\relax
\mciteBstWouldAddEndPuncttrue
\mciteSetBstMidEndSepPunct{\mcitedefaultmidpunct}
{\mcitedefaultendpunct}{\mcitedefaultseppunct}\relax
\EndOfBibitem
\bibitem[Sahoo \latin{et~al.}(2022)Sahoo, Theeyancheri, and
  Chakrabarti]{Sahoo2022}
Sahoo,~R.; Theeyancheri,~L.; Chakrabarti,~R. Transport of a self-propelled
  tracer through a hairy cylindrical channel: interplay of stickiness and
  activity. \emph{Soft Matter} \textbf{2022}, \emph{18}, 1310--1318\relax
\mciteBstWouldAddEndPuncttrue
\mciteSetBstMidEndSepPunct{\mcitedefaultmidpunct}
{\mcitedefaultendpunct}{\mcitedefaultseppunct}\relax
\EndOfBibitem
\bibitem[Lu \latin{et~al.}(2022)Lu, Huang, and Luo]{Lu2022}
Lu,~R.-X.; Huang,~J.-H.; Luo,~M.-B. A simulation study on the subdiffusion of
  polymer chains in crowded environments containing nanoparticles. \emph{Phys.
  Chem. Chem. Phys.} \textbf{2022}, \emph{24}, 3078--3085\relax
\mciteBstWouldAddEndPuncttrue
\mciteSetBstMidEndSepPunct{\mcitedefaultmidpunct}
{\mcitedefaultendpunct}{\mcitedefaultseppunct}\relax
\EndOfBibitem
\end{mcitethebibliography}
\end{document}